\documentclass[
aps,%
11pt,%
final,%
notitlepage,%
oneside,%
twocolumn,%
nobibnotes,%
nofootinbib,%
superscriptaddress,%
noshowpacs,%
centertags]%
{revtex4}
\begin{document}
%\selectlanguage{russian} %
\selectlanguage{english} %
\title{The KHOLOD Experiment: A Search for a New Population of Radio Sources}
\author{\firstname{Yu.N.}~\surname{Parijskij}}
\affiliation{Special Astrophysical Observatory, Russian Academy of Sciences,
Nizhnii Arkhyz, Karachaev-Cherkessian Republic, Russia}
\author{\firstname{N.S.}~\surname{Soboleva}}
\affiliation{St. Petersburg Branch of the Special Astrophysical
Observatory, Russian Academy of Sciences, St. Petersburg, Russia}
\author{\firstname{A.V.}~\surname{Temirova}}
\affiliation{St. Petersburg Branch of the Special Astrophysical
Observatory, Russian Academy of Sciences, St. Petersburg, Russia}
\email[E-mail:]{adelina_temirova@mail.ru}
\author{\firstname{N.N.}~\surname{Bursov}}
\affiliation{Special Astrophysical Observatory, Russian Academy of
Sciences, Nizhnii Arkhyz, Karachaev-Cherkessian Republic, Russia}
\author{\firstname{O.P.}~\surname{Zhelenkova}}
\affiliation{Special Astrophysical Observatory, Russian Academy of
Sciences, Nizhnii Arkhyz, Karachaev-Cherkessian Republic, Russia}
\begin{abstract}

Published data from long-term observations of a strip of sky at declination
$\delta\sim5^{\circ}$ carried out at 7.6\,cm on the \mbox{RATAN-600} radio
telescope are used to estimate some statistical properties of radio sources.
Limits on the sensitivity of the survey due to noise imposed by background
sources, which dominates the radiometer sensitivity, are refined. The vast
majority of noise due to background sources is associated with known radio
sources (for example, from the NVSS with a detection threshold of 2.3\,mJy)
with normal steep spectra ($\alpha = 0.7-0.8, S\propto\nu^{-\alpha}$), which
have also been detected in new deep surveys at decimeter wavelengths. When
all such objects are removed from the observational data, this leaves another
noise component that is observed to be roughly identical in independent
groups of observations. We suggest this represents a new population of radio
sources that are not present in known catalogs at the 0.6\,mJy level at
7.6\,cm.
The studied redshift dependence of the number of steep-spectrum
objects shows that the sensitivity of our survey is sufficient to detect
powerful FRII radio sources at any redshift, right to the epoch of formation
of the first galaxies. The inferred new population is most likely
associated with low-luminosity objects at redshifts $z<1$. In spite of the
appearance of new means of carrying out direct studies of distant galaxies,
searches for objects with very high redshifts among steep and ultra-steep
spectrum radio sources remains an effective method for studying the early
Universe.

\end{abstract}
\maketitle
\section{Introduction}

%
%%++ Figure:1
\begin{figure*}
\begin{center}
\onelinecaptionsfalse
\centerline{
\includegraphics[angle=0,width=0.6\textwidth,clip]{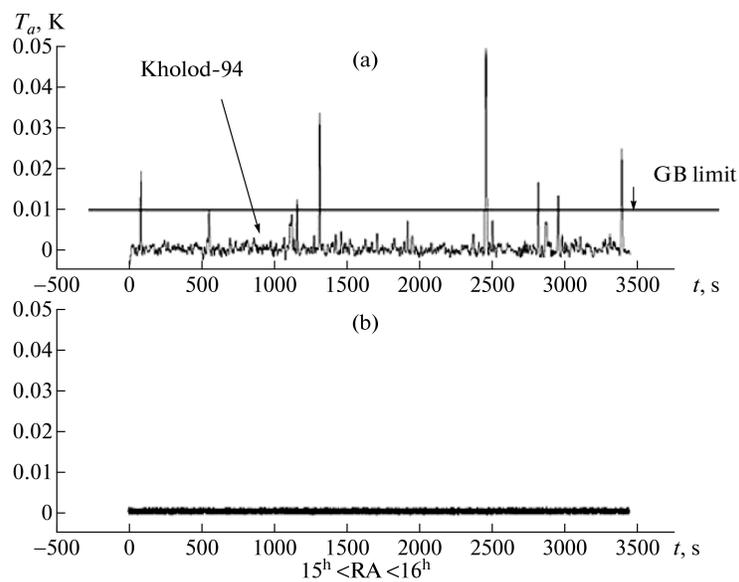}
}
\setcaptionmargin{0mm}
\captionstyle{normal}
\caption{
(a) Example of real noise taking into account background sources (the
threshold for the GB6 catalog is shown). (b) Expected radiometer noise at
the central wavelength of 7.6\,cm for the KHOLOD experiment for monthly
averaging of the data.
}
\label{fig1}
\end{center}
\end{figure*}
%
%%++ Figure:2
\begin{figure*}
\begin{center}
\onelinecaptionsfalse
\centerline{
\includegraphics[angle=0,width=0.6\textwidth,clip]{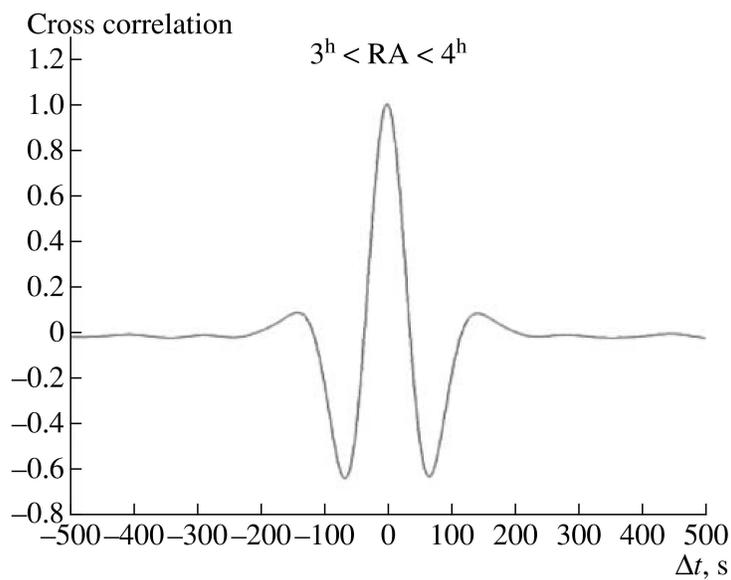}
}
\setcaptionmargin{0mm}
\captionstyle{normal}
\caption{
Cross correlation of two independent groups of observations in the KHOLOD-94
experiment ($\lambda=7.6\,cm$). The horizontal axis plots the time shift
$\Delta t$ between these groups of observations.
}
\label{fig2}
\end{center}
\end{figure*}
The RATAN-600 radio telescope was initially proposed as a supplementary
instrument to ordinary reflecting radio telescopes (paraboloids), in the form
of a multi-frequency transit telescope suitable for surveys, with a fairly
large daily field of view and resolution higher than a paraboloid of the same
area. There are virtually no survey instruments operating at centimeter
wavelengths. The reason is essentially that the vast majority of background
radio sources have spectra that fall off roughly linearly with decreasing
wavelength. The field of view of a radio telescope falls off in proportion
to the square of the wavelength; the time for a radio source to pass through
the telescope beam also falls off with decreasing wavelength. All this means
that nearly four orders of magnitude more time is required for a blind survey
at centimeter than at decimeter wavelengths, of the same area of the sky and
to the same depth in terms of background radio sources\,\cite{Ekers}.
%
%%++ Figure:3
\begin{figure*}
\begin{center}
\onelinecaptionsfalse
\setcaptionmargin{0mm}
\captionstyle{normal}
\centerline{
\includegraphics[angle=0,width=0.6\textwidth,clip]{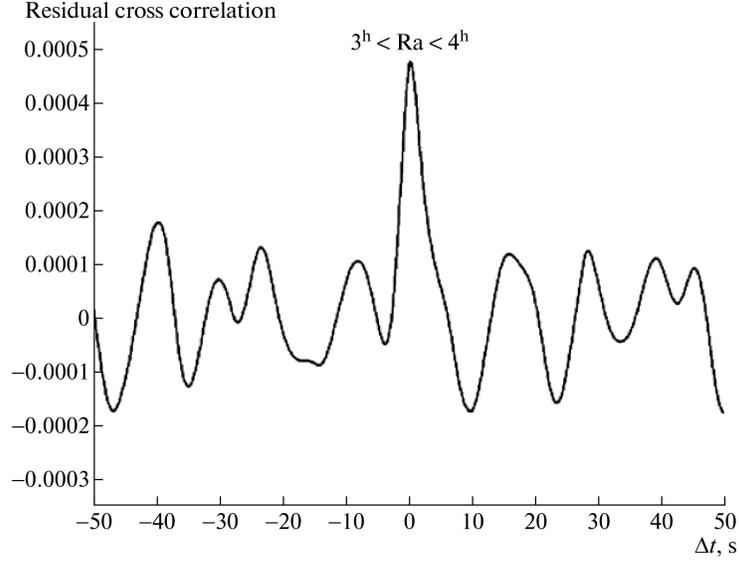}
}
\caption{
Noise due to background sources with flux densities less than 0.6\,mJy for
NVSS objects with normal steep spectra, which are clearly visible in the
cross correlation of two independent sets of data, after filtration of the
noise from all known NVSS sources.
}
\label{fig3}
\end{center}
\end{figure*}
Deep surveys of large regions of the sky were begun on the RATAN-600 by radio
astronomers of the Sternberg Astronomical Institute very soon after the
completion of its construction. The short time for the passage of sources
through the radio-telescope beam limited the sensitivity of the survey.
The hope that sources with inverted spectra would dominate the
centimeter-wavelength sky did not prove justified. The creation by the
group of D.V.~Korol'kov of a new generation of cryoradiometers with
sensitivities appreciably higher than previous such instruments\,\cite{par1},
as well as the decision not to try to survey large areas of the sky, was
expected to enhance the depth of blind surveys.
%
%%++ Figure:4 and 5
\begin{figure*}
\setcaptionmargin{0mm}
\captionstyle{normal}
\centerline{
\includegraphics[angle=0,width=0.45\textwidth,clip]{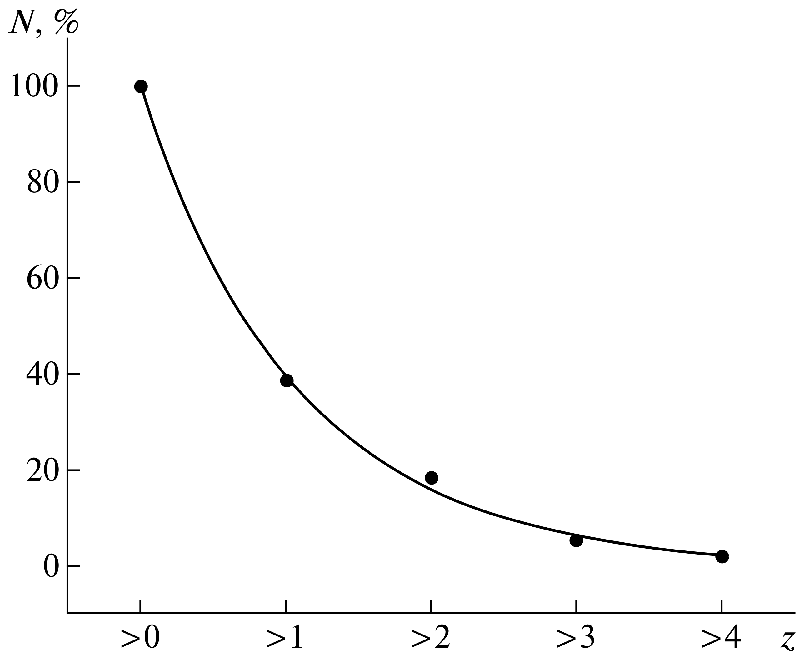}
\includegraphics[angle=0,width=0.45\textwidth,clip]{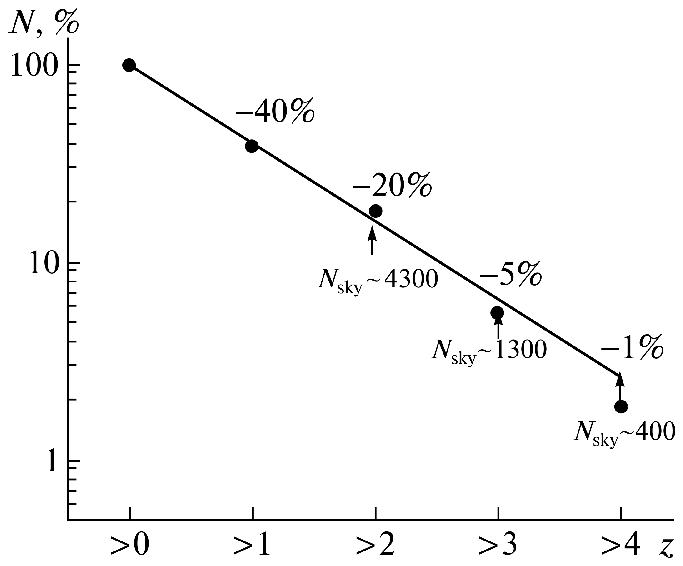}
}
\caption{
Left - drop in the number of radio sources with redshift according to data from
the Big Trio project~\cite{par14}.
Right -
estimate of the expected number of radio sources in the Universe with steep
spectra as a function of redshift, based on the Big Trio data.
}
\label{fig4}
\end{figure*}
The appearance in the past 10 years of
deep surveys a factor of 100 more sensitive than previous surveys cardinally
changes the strategy used for blind surveys. Such surveys on the RATAN-600
telescope now yield virtually no new radio sources at centimeter wavelengths
that are not present in the NVSS and FIRST catalogs at 21 cm. The only new
radio sources that could appear in RATAN-600 observations but be absent from
the NVSS catalog are objects whose maximum radiation occurs at centimeter
wavelengths (for example, whose spectra display synchrotron self-absorption).
There are few such objects, and for blind surveys of such objects, it is more
effective to expand the survey area, rather than increase the depth of
observations of a small survey region. Repeating observations of a single
region increases the sensitivity as the square root of the number of
observations, while the dependence of the number of objects on flux density
is nearly inear ($N\sim S^{-1}$) at centimeter wavelengths. Therefore, for
a fixed total observing time, expanding the survey area provides
a substantial gain in the number of new objects, compared to carrying out
repeated observations of a single limited region.

The KHOLOD experiment
(1980-2000) combined a survey of discrete radio sources with very
deep studies of the anisotropy of the cosmic microwave
background\,\cite{par2,par3,vo1,par4,par5}. The sensitivity to the population
of radio sources with normal steep spectra in this experiment was appreciably
higher than the sensitivity of catalogs available at the beginning of the 1980s,
and the vast majority of the objects detected were new.
Only with the appearance of the Texas UTRAO new-generation 80-cm
catalog\footnote{ Kindly presented to us by Prof.~J.~Douglas prior to its
publication.} did it become possible to obtain a fairly deep sample of
objects with steep and ultra-steep spectra. The curvature of the radio
spectrum is one of the main indicators that can be used to identify candidate
distant objects\,\mbox{\cite{Kapahi,McCarthy,Pedani}}.

The ``Big Trio'' project,
with the participation of three major facilities (RATAN-600, the VLA, and the
6-m telescope of the Special Astrophysical Observatory (SAO) of the Russian
Academy of Sciences) \cite{vo1,Goss1,bur1,Goss2,kop1,kop2,par6,Fletcher,
par7,par8,par9,vo2,dod1,Pursimo,par10,bur2,par11,sob1,par12,vo3,par13,vo4,
sob2,vo5,vo6,sob3,kop3,kop4,par14}, was aimed at using this sample of sources
to select candidate very distant galaxies. The sensitivity of the RATAN-600
was sufficient to detect powerful radio sources at all redshifts in blind
surveys.

All these problems led to a simpler approach -- observing radio sources at
centimeter and millimeter wavelengths with prolonged integration times,
pointing a radio telescope at an NVSS or FIRST object. Unfortunately, even
this method does not enable the construction of spectra of all the objects
detected at decimeter wavelengths: sensitivities of several microjansky are
required, which is currently achievable only in very deep VLA surveys of
very small areas of sky\,\cite{Fomalont}. Blind surveys of such small areas
provide important information about the properties of low luminosity radio
sources and, as a rule, no new information about classical high-luminosity
radio galaxies (FRI, FRII, QSR).

\section{Noise due to background sources}

In the 1950s, the sensitivity of meter-wavelength radio telescopes was
limited by noise due to background sources, and the hope was that this noise
would be much smaller at centimeter wavelengths (proportional to the
wavelength cubed\,\cite{Khaikin}). Now, virtually the entire radio astronomy
range is subject to noise due to background sources for observations with
ordinary reflecting telescopes. Observations on the southern sector of
RATAN-600 using its flat reflector encounter the same background-source noise
limit as a paraboloid with the same surface area. In observations at medium
and high hour angles using a single sector, the effective antenna beam is
appreciably smaller than a paraboloid of the same area, and there is
consequently less ``confusion'' noise.

Two-dimensional mapping at high hour angles applying reasonable data-reduction
techniques enables reduction of the confusion noise by nearly an order of
magnitude\,\cite{Chepurnov}. Even in one-dimensional surveys, optimal
filtration of the data yields almost the same results as filtration using
two-dimensional maps. Even simple, ordinary, close two-beam scanning (the
first derivative) substantially reduces the confusion noise. In the 1950s,
Bracewell\,\cite{Brac1,Brac2} proposed to use the second derivative
(``Cord Construction in Radio Astronomy'') to reduce the confusion noise.

This method proved to be very effective for the RATAN-600 beam at high hour
angles. It leaves only noise from point-like radio sources that fall in
a central strip of the survey half a beam wide in declination. However,
with new-generation cryoradiometers, the confusion noise dominates
the radiometer noise, even in modest cycles of observations.
This is clear in the KHOLOD experiment: the expected radiometer noise at
7.6\,cm for a monthly observing cycle is appreciably lower than the real
noise (Fig.~\ref{fig1}), and the correlation between the data for two small
independent observing cycles is nearly 100\%. Whereas all the sources
detected were new in the first observations for the KHOLOD experiment in
1980, with the compilation of catalogs such as NVSS and FIRST, the detected
objects now have counterparts in these deep surveys. The real confusion noise
is clearly visible in a cross correlation of two independent groups of
observations (Fig.~\ref{fig2}): the shape of the curve corresponds fully to
theoretical expectations in the absence of independent noise in the two sets
of data.
Confusion noise dominates, in spite of the filtration of noise due to
background sources that are far from the beam axis using the ``second
difference or derivative'', now called the Mexican Hat
method\,\cite{Brac1,Vielva}.

When the model (the NVSS objects extrapolated to 7.6\,cm using a typical
steep spectrum smoothed with the RATAN-600 beam) and observations are
processed using a Mexican Hat filter to filter out background sources far
from the beam axis, they display a high correlation. The Mexican Hat filter
appreciably lowers the confusion noise, but there remains the high noise of
the radiometer. To estimate the contribution to the confusion noise by a new
population of sources that are not present in known catalogs, we removed
from the 7.6-cm data all NVSS objects with flux densities close to their
limiting value ($S_{lim}=2.3\,mJy$ at 21\,cm, which corresponds to
$S_{lim}=0.6\,mJy$ at 7.6\,cm, assuming a typical steep spectrum).

Figure~\ref{fig3} presents the residual cross correlation (in arbitrary
units) after removing all known features brighter than 0.6\,mJy. Thus, all
features distinguishable as individual sources can be identified with NVSS
sources. There remains noise due to weak background sources at the sub-mJy
level at 7.6\,cm, which is fairly firmly detected statistically (via the
cross correlation of independent data sets) in deep surveys such as
KHOLOD (see also the circum-zenith survey described in \cite{bur3,sem1}).
These sources are not detected in the NVSS catalog at 21 cm.

\section{The nature of the weak radio sources}

%
%%++ Figure:6
\begin{figure*}
\begin{center}
\onelinecaptionsfalse
\setcaptionmargin{0mm}
\captionstyle{normal}
\centerline{
\includegraphics[angle=0,width=0.6\textwidth,clip]{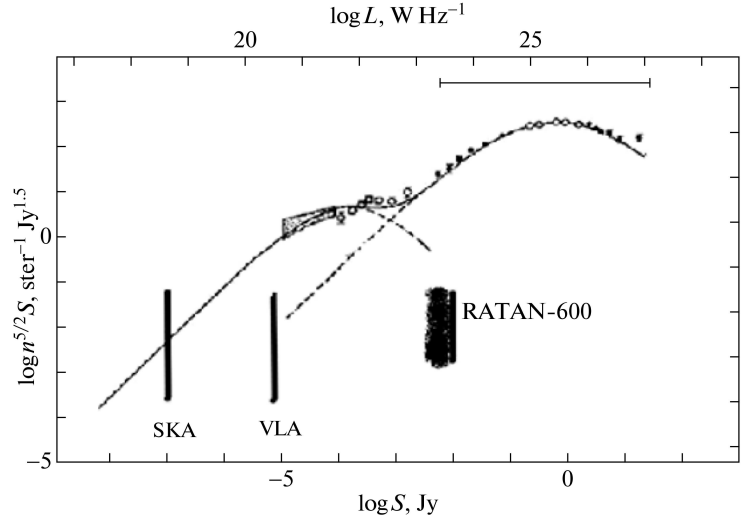}
}
\caption{
Estimate of the penetrating power of radio telescopes of various classes for
studies of galaxies with various luminosities (dashed). The upper horizontal
line marks the range occupied by flux densities of classical radio galaxies
in the redshift range $0<z<8$, which encompasses the entire Universe to the
epoch of reionization. The vertical lines show the limits attainable on the
RATAN-600 and VLA and the expected limit for the Square Kilometer Array.
}
\label{fig6}
\end{center}
\end{figure*}
We show below that this new population of radio sources visible statistically
in deep \mbox{RATAN-600} surveys is most likely associated with low-luminosity
objects. Useful nformation about the population of objects with high radio
luminosities can be extracted from the Big Trio project. Photometric and
spectral redshifts of objects with steep and ultra-steep spectra in the field
of view of the KHOLOD experiment are presented in \cite{par14}.
Redshifts could be determined for 54 of the 71 studied objects: of these ten
have $z>2$, three have $z>3$, and one has $z>4$ (RCJ~0311+0507, $z=4.514$).
%
%%++ Figure:7
\begin{figure*}
\onelinecaptionsfalse
\setcaptionmargin{0mm}
\captionstyle{normal}
\centerline{
\includegraphics[angle=0,width=0.45\textwidth,clip]{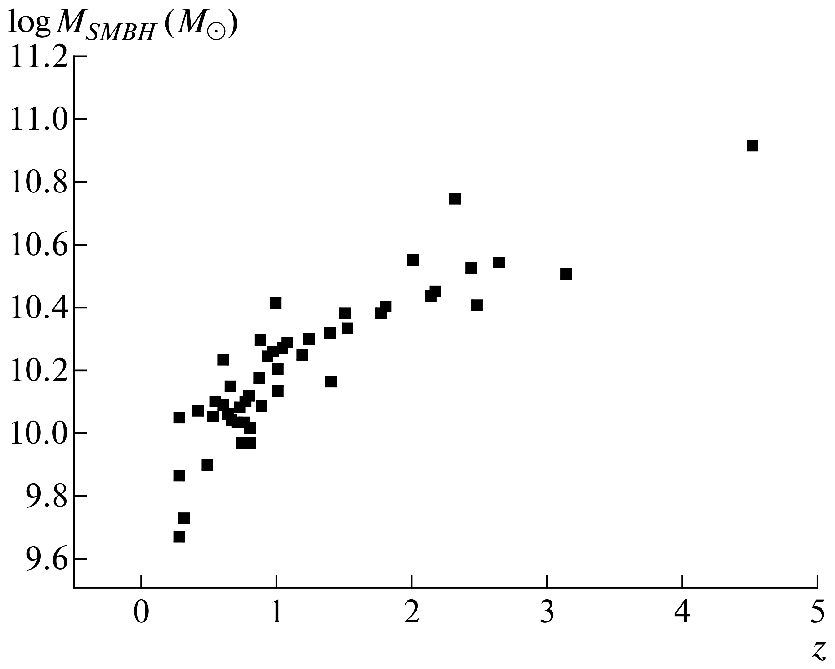}
\includegraphics[angle=0,width=0.4\textwidth,clip]{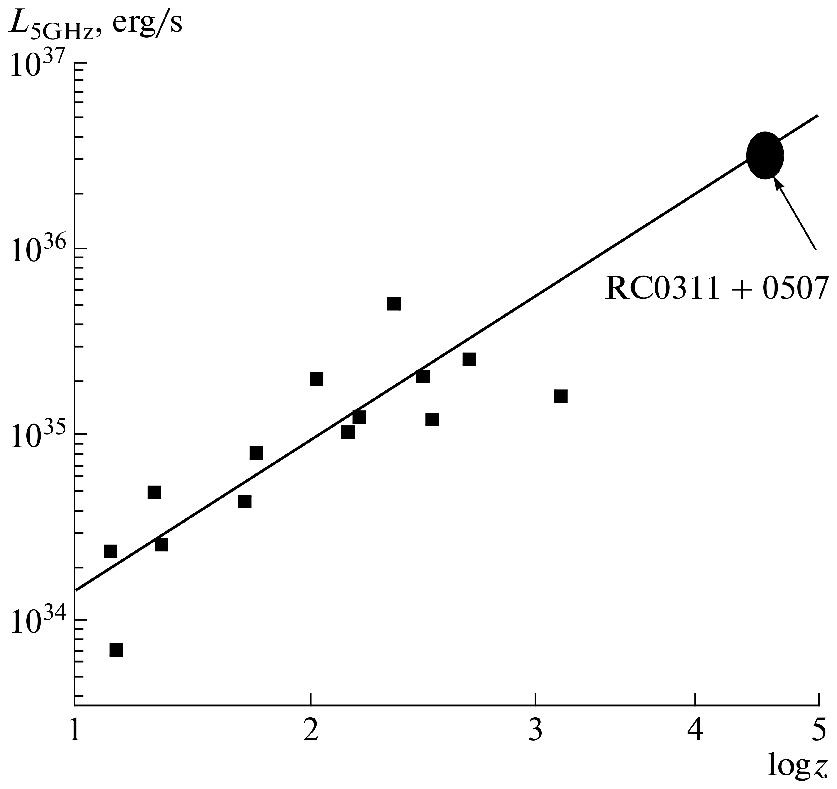}
}
\caption{
Left -- masses of supermassive black holes detected in the KHOLOD field of view as a
function of redshift for the population of steep and ultra-steep spectrum
objects. The higher the redshift, the more massive the black holes required
to explain their radio luminosities.
Right --
redshift dependence of the absolute 5-GHz luminosity for high-redshift FRII
objects, based on the collected data of Miley and de Breuck~\cite{Miley}.
The radio luminosity is proportional to $M^{2.5}_{SMBH}$, where $M_{SMBH}$ is
the mass of the supermassive black hole~\cite{Salucci}. An object from
the Big Trio project (RC~J0311+0507) has the most massive black hole,
according to \cite{par14}.
}
\label{fig7}
\end{figure*}
Figure~\ref{fig4} (left) shows the drop in the number of radio sources with
redshift for these 54 objects. Based on data published in
2007-2010~\cite{par14,Miley,Jarvis,Ishwara,Vardoulaki,Bornancini}, there are
currently 194 known radio sources with $z>2$, 47 with $z>3$, and 9 with $z>4$.
It is possible to extrapolate these data to the entire sky, taking into
account the KHOLOD survey region (0.005 ster; Fig.~\ref{fig4}, right). Note that all
powerful radio galaxies in the  KHOLOD field of view were detected at the
limiting flux level of the RC~catalog of 5-10\,mJy (the Big Trio project),
right to the epoch of secondary reionization ($z=10$). This can be seen in
Fig.~\ref{fig6} (which shows a normalized curve for the radio-source counts
at 21\,cm~\cite{Condon}).

Powerful radio galaxies are believed to be associated with the activity of
supermassive black holes at the centers of their giant elliptical host
galaxies. The Big Trio data can be used to estimate the redshift distribution
of such galaxies in the Universe (Fig.~\ref{fig7}, left). The Big Trio data are in
agreement with recently published international data for radio sources
with high redshifts (Fig.~\ref{fig7}, right).

According to RATAN-600 data, the surface density of objects with high
redshifts is much higher than the value detected in surveys of the entire
sky, due to the lower limiting flux density of the KHOLOD experiment. It is
important here that the statistical confusion noise is approximately equal
to the flux density of radio sources whose surface density is about one per
telescope beam\,\cite{Esepkina}. In the KHOLOD survey, the effective beam size
at 7.6\,cm was about three square minutes of arc, which gives more than
a million beams on the sky.
%
%%++ Figure:9
\begin{figure*}
\begin{center}
\onelinecaptionsfalse
\setcaptionmargin{0mm}
\captionstyle{normal}
\centerline{
\includegraphics[angle=0,width=0.5\textwidth,clip]{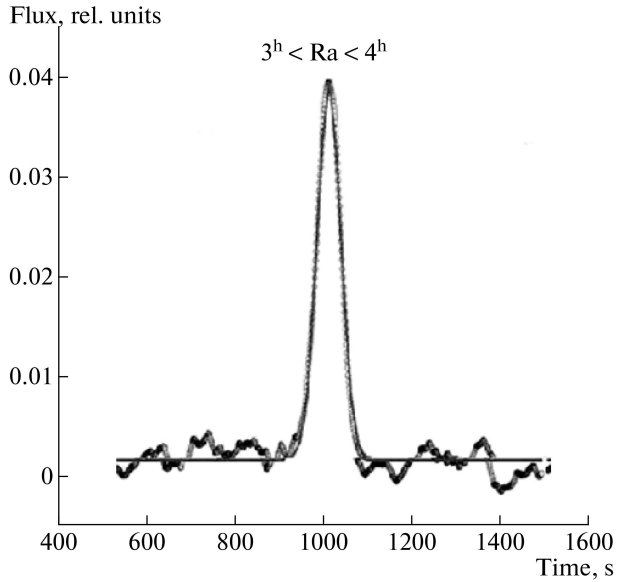}
}
\caption{
Individual scan of the radio source RC~J0311+0507 at 7.6\,cm (RATAN-600, 1980).
}
\label{fig9}
\end{center}
\end{figure*}
%
%%++ Figure:10
\begin{figure*}
\begin{center}
\onelinecaptionsfalse
\setcaptionmargin{0mm}
\captionstyle{normal}
\centerline{
\includegraphics[angle=0,width=0.7\textwidth,clip]{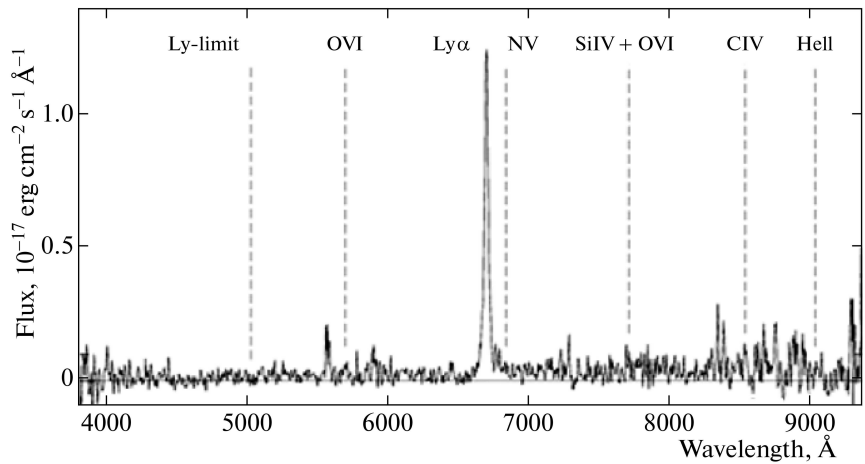}
}
\caption{
Optical spectrum of the host galaxy of RCJ 0311+0507 (2004) obtained using the
SCORPIO spectrograph\,\cite{afan} mounted on the 6-m telescope of the Special
Astrophysical Observatory.
}
\label{fig10}
\end{center}
\end{figure*}
The dependence of the number of FRII radio galaxies on the survey sensitivity
estimated in \cite{Garrett} shows that, even with a sensitivity of 30 nJy at
21\,cm, the surface density remains at about 50 sources per square degree,
independent of the survey depth in the range 1\,mJy-30\,nJy. The number of
antenna beams per square degree in the KHOLOD experiment (and
the circum-zenith survey) is hundreds of times greater, supporting our
conclusion concerning the low luminosities of objects in the
``new population'' of centimeter-wavelength radio sources. Their luminosities
could be close to those of active spiral galaxies.

\section{Conclusion (the Big Trio project and the $21^{st}$ century)}

The selection of candidate distant galaxies from among radio sources with
ultra-steep spectra is an effective method for studying the early Universe
in the $21^{st}$ century. The vast majority of galaxies with $z>1$ have been found
using this method, although it remains unclear why objects with ultra-steep
spectra have high radio luminosities. Powerful radio galaxies are accessible
to observations with existing instruments at any redshift, right to the epoch
of galaxy formation immediately after the end of the Dark Age $(z\sim10)$.
Selecting ultra-steep-spectrum sources in radio surveys and following this up
with optical studies of their host galaxies makes it possible to determine
their redshifts and other properties. The Big Trio project was one of
the first to use this approach. The penetrating ability of the RATAN-600 and
SAO 6-m telescope are sufficient to confidently detect the population of
FRII radio galaxies at high redshifts in the radio and measure the spectral
properties of their host galaxies, as is illustrated by the example of the
radio galaxy RC~J0311+0507 (Figs.~\ref{fig9},~\ref{fig10}).

The signal-to-noise ratio in this case is \mbox{$S/N>100$}, the object's flux
densities are \mbox{$S_{7.6\,cm}=135\,mJy$} and \mbox{$S_{21\,cm}=537\,mJy$}, its the
largest angular size in the radio is 2.8, and its spectral index is
$\alpha=1.37$~\cite{Goss2,bur2,par11,kop3,par14}. Instrumental progress in
all spectral ranges enables refinement of the role of this approach in
current and future studies. Successful attempts to search for
first-generation galaxies in deep optical and infrared ground-based
and space-based observations have already been made. The use of gravitational
lensing makes it possible to find candidate galaxies with redshifts $z=7-9$,
independent of their radio properties~\cite{Lehnert}. Nevertheless, we
believe that behind searches for powerful radio sources lies the need to
select galaxies with powerful active nuclei. It is current believed that
these could be indicators of the earliest clusters. The radio luminosities
of galaxies are associated with activity of the supermassive black holes in
their nuclei. It unexpected was found (in the Big Trio project) that,
the higher the redshift, the higher the inferred black-hole masses. Moreover,
theory predicts that the radio luminosity should depend on both the accretion
rate and the rotational velocity of the black hole. It is believed that
angular momentum can be transferred in collisions between galaxies or black
holes. The accessibility of powerful FRII radio galaxies at any distance in
the Universe to observations even with radio telescopes with only moderate
sensitivities and their connection with first-generation galaxies mean that
these objects can be used to estimate deviations from the standard cosmology,
by comparing the ages of the stellar populations of the host galaxies derived
from multi-color photometry and the age of the Universe at the corresponding
redshifts. Thus, radio and optical (infrared) selection supplement each other.

\acknowledgements{
This work was partially supported by the Russian Foundation for Basic
Research (projects 11-02-00489a, 11-02-12036-ofi-m). The authors thank
O.V.~Verkhodanov for the use of his program $fgr$.
}

\end{document}